\documentclass[aps,twocolumn,amssymb,amsmath,floats,showpacs,pre]{revtex4-1}
\usepackage{bm}
\usepackage{graphicx}
\usepackage{xcolor}
\usepackage[toc,page]{appendix}
\usepackage{comment}
\usepackage[unicode=true]{hyperref}
\usepackage{enumerate}
\usepackage{dsfont}
\usepackage{amsmath}
\usepackage{soul}

\usepackage[normalem]{ulem}
\usepackage[ruled,vlined]{algorithm2e}
\SetKwProg{Fn}{Function}{}{}
\SetKwInOut{Input}{Input}
\SetKwInOut{Output}{Output}
\SetKwInOut{Parameter}{Parameter}
\usepackage{multirow} 
\usepackage{lineno} 

\graphicspath{{images/}}
\begin{document}

\title{Activity Regeneration from Silent States in Neuronal Networks with Transient Synaptic Memory}

%

\author{Mozhgan Khanjanianpak$^{1}$}
\email{Corresponding author: khanjanianpak@gmail.com}
\author{Alireza Valizadeh$^{2,3}$}

\vspace{0.9cm} 

\affiliation{$^{1}$ \mbox{Pasargad Institute for Advanced Innovative Solutions (PIAIS), Tehran 1991633357, Iran}}
\affiliation{$^{2}$ \mbox{Zapata Briceño Institute for Human Intelligence Research, Madrid, Spain}}
\affiliation{$^{3}$ \mbox{Physics Department, Institute for Advanced Studies in Basic Sciences (IASBS), Zanjan 45137-66731, Iran}}


\begin{abstract}
	Transient synaptic memory has emerged as a potential mechanism for maintaining short-term information even in the absence of persistent neuronal activity. However, it remains unclear whether the hidden synaptic state alone contains sufficient information to predict the future evolution of neuronal networks after activity has ceased. Here, we introduce a minimal neuronal network model with finite-lifetime synapses and investigate the mechanism underlying spontaneous activity regeneration following complete neuronal silence. We show that the residual synaptic configuration at the first silent state already determines whether network activity terminates after a single activation cycle or spontaneously regenerates an additional cycle. By analyzing this synaptic-memory snapshot, we identify the Latent Excitatory Recruitment (LER) capacity, quantified by the cumulative number of fresh excitatory neurons, as a near-perfect predictor of multi-cycle dynamics without continuing the subsequent network simulation. Remarkably, these distinct dynamical outcomes emerge in an otherwise homogeneous neuronal network, demonstrating that transient synaptic memory alone is sufficient to generate diverse future dynamics. Our findings provide a mechanistic explanation for activity regeneration from a residual synaptic state and suggest that short-term memory is encoded not only in ongoing neuronal activity but also in the latent synaptic configuration that preserves the network's capacity to recruit new neuronal assemblies. More broadly, the proposed snapshot-based framework offers a new perspective for predicting and potentially controlling the future evolution of neuronal networks.
\end{abstract}

\maketitle

\section{Introduction}
\label{sec:introduction}
The ability to temporarily retain and manipulate information is one of the fundamental functions of the brain, supporting perception, decision making, reasoning, and goal-directed behavior \cite{baddeley2012working, d2015cognitive, van2023turning}. This capability, commonly referred to as working memory \cite{GOLDMANRAKIC1995477, barbosa2020interplay}, has traditionally been thought to rely on persistent neuronal activity that remains elevated during the delay period following stimulus disappearance \cite{wang2001synaptic,  constantinidis2018persistent, curtis2021persistent}. Experimental studies together with computational models have demonstrated that recurrent excitation can sustain stable patterns of activity over behaviorally relevant time scales, providing a plausible neural substrate for short-term memory \cite{amit1989modeling,  fuster1971neuron, funahashi1989mnemonic}. Consequently, persistent activity has long been regarded as the canonical mechanism underlying the temporary maintenance of working-memory representations in cortical circuits \cite{riley2016role}.

However, accumulating experimental and theoretical evidence suggests that persistent neuronal firing alone is insufficient to fully account for the maintenance of short-term memories \cite{stokes2015activity, lundqvist2018working, miller2018working, rose2016reactivation}. Accumulating neurophysiological evidence has shown that memory-related information remains recoverable even during periods of weak or apparently absent delay-period activity, suggesting that working memory can be maintained in hidden network states rather than continuous persistent firing \cite{rose2016reactivation, wolff2017dynamic, stokes2015activity, masse2019circuit}. These observations have led to the concept of \emph{activity-silent} working memory, in which transient synaptic modifications temporarily preserve information and can subsequently reactivate the corresponding neuronal representation when the network state becomes favorable for reactivation \cite{stokes2015activity, mongillo2008synaptic, rose2016reactivation, wolff2017dynamic}. Rather than replacing the classical persistent-activity framework, this emerging view suggests that persistent neuronal activity and transient synaptic states represent complementary mechanisms that jointly support the maintenance of working-memory representations \cite{miller2018working, lundqvist2018working, stokes2015activity, masse2019circuit}.

Despite the growing evidence supporting activity-silent memory, one fundamental question remains unanswered: if transient synaptic states preserve information after neuronal activity has completely disappeared, what determines whether this hidden memory remains latent or reactivates the network to generate a new activity cycle? More specifically, is the residual synaptic-memory state itself sufficient to determine the future evolution of the network, distinguishing between permanent termination and spontaneous activity regeneration? This question is particularly intriguing because the network considered here is structurally homogeneous, suggesting that any diversity in its future evolution must arise from its transient synaptic state rather than from structural heterogeneity. 

Existing studies have primarily investigated how transient synaptic dynamics preserve and reactivate memory representations \cite{mongillo2008synaptic, stokes2015activity, masse2019circuit, kaminski2020combined}, whereas comparatively little attention has been paid to identifying quantitative signatures within the residual synaptic state that predict the subsequent evolution of network dynamics. Identifying such predictors would not only deepen our mechanistic understanding of short-term memory maintenance, but also address a more fundamental question: whether the future evolution of a neuronal network is already encoded in a single synaptic-memory snapshot.

Answering this question requires shifting the focus from neuronal activity to the residual synaptic configuration itself. During periods of complete neuronal silence, the instantaneous activity pattern no longer contains information capable of distinguishing the network's future evolution. In contrast, synapses with finite lifetimes continue to evolve, preserving a transient trace of preceding activity. The residual synaptic configuration therefore constitutes an evolving dynamical substrate rather than merely a passive remnant of previous activation. If the information governing future activity is indeed embedded within this latent synaptic configuration, then predicting network evolution becomes a problem of characterizing the network's latent reactivation capacity encoded in a single synaptic-memory snapshot, without requiring continued simulation of the full neuronal dynamics.

To address these questions, we introduce a minimal neuronal network model in which memory is represented by synapses with finite lifetimes, allowing transient synaptic traces to persist even after neuronal activity has completely disappeared. We then investigate how the residual synaptic configuration present during the first silent period shapes the subsequent evolution of network dynamics. By distinguishing between two dynamical classes of neurons, termed \emph{ghost} and \emph{fresh} neurons, we identify a simple quantitative measure of the network's latent excitatory reactivation capacity encoded in a single synaptic-memory snapshot. Remarkably, this quantity accurately predicts whether network activity terminates after a single activation cycle or spontaneously regenerates an additional cycle. These findings demonstrate that the future evolution of the system is already encoded in the residual transient synaptic state and can therefore be inferred directly from a single synaptic-memory snapshot, without continuing the subsequent neuronal dynamics.

The remainder of this paper is organized as follows. Section~\ref{sec:model} introduces the proposed neuronal network model, its transient synaptic-memory dynamics, and the observables used throughout the study. Section~\ref{sec:results} presents the numerical results, including the characterization of the observed dynamical regimes, the identification of a snapshot-based predictor of activity regeneration, and an analysis of its robustness under different spontaneous activation rates. In Section~\ref{sec:discussion}, we discuss the implications of the findings in the context of contemporary theories of short-term memory and dynamic coding, outlines the limitations of the present work, and highlights several directions for future research. Finally, the paper is concluded in Section~\ref{sec:conclusion}

\section{Model Description} 
\label{sec:model}
Conventional neural-network models usually describe network dynamics solely in terms of neuronal activity. In threshold-based networks, the state of the system at time $t$ is typically represented by a binary activity vector indicating which neurons are active. Synaptic interactions determine how activity propagates from one time step to the next, but the synapses themselves are often treated as static entities whose effect is instantaneous.

Although this approximation has been highly successful and forms the basis of many influential models in theoretical neuroscience, it neglects an important biological feature: synaptic effects persist in time. In real neuronal systems, activation of a synapse does not produce an instantaneous interaction that disappears immediately. Neurotransmitter release, receptor activation, and postsynaptic currents may remain effective over finite time intervals whose duration depends on the underlying synaptic mechanisms. Moreover, excitatory and inhibitory synapses generally exhibit different temporal characteristics.

We introduce a synaptic-lifetime network (SLN) model, a discrete-time excitatory–inhibitory neural network in which synaptic connections remain active for a finite lifetime after presynaptic activation. The SLN model was originally introduced in algorithmic form in our previous work \cite{khanjanianpak2024emergence}, where it was shown that finite synaptic activity times can generate rich oscillatory phenomena, including transient cycles, bursts of activity states, and complex temporal patterns. However, the model lacked a compact mathematical formulation. Here we reformulate it as a discrete-time dynamical system and identify the minimal state variables governing its evolution.

The key conceptual difference from conventional threshold networks is that the state of the system is not determined solely by neuronal activity. Instead, the network possesses an explicit synaptic state variable that stores information about previous activation events.

\subsection{State Variables}
\begin{figure*}[t!]
	\centering    \includegraphics[width=1\linewidth]{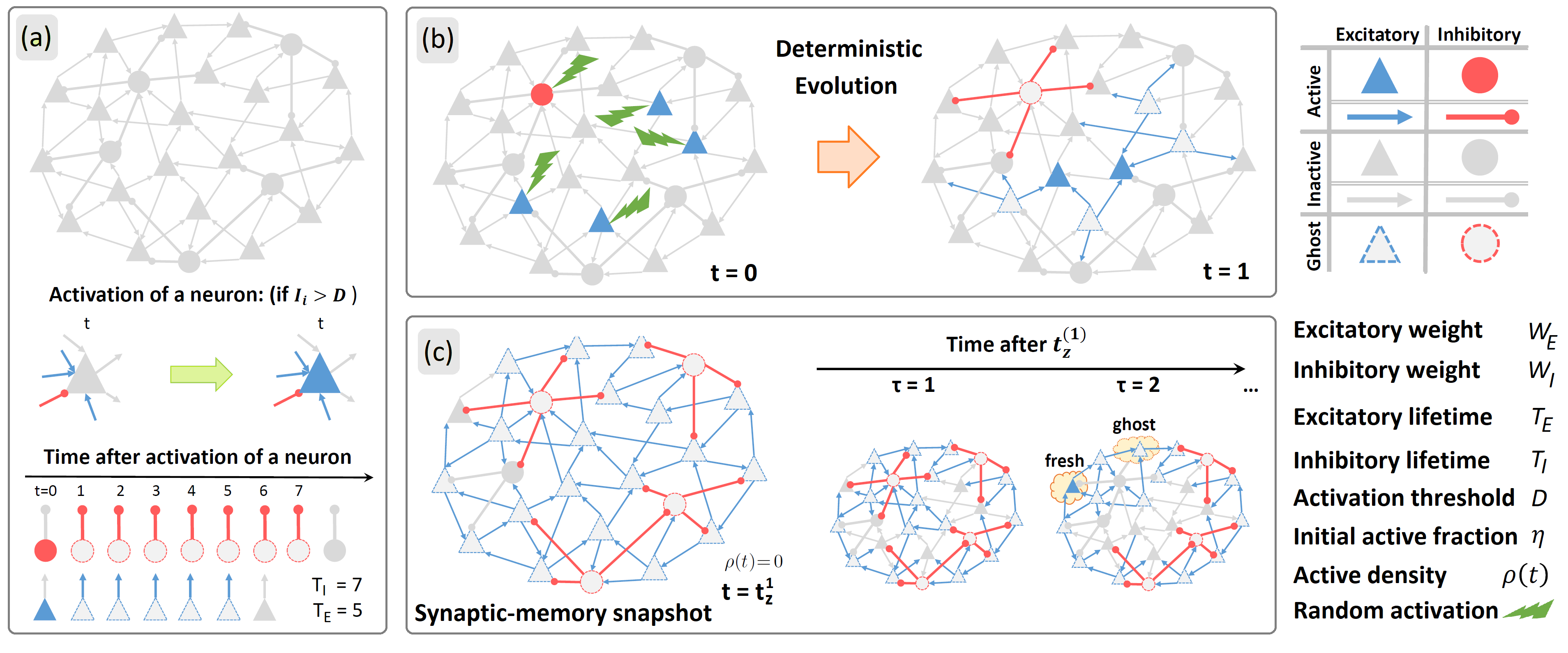}
	\caption{\textbf{Schematic illustration of the proposed neuronal network model.} 	(a) General dynamics of the model. A neuron becomes active when its total synaptic input exceeds the activation threshold $D$. One time step later, its outgoing synapses become active and remain active for a finite lifetime ($T_E$ for excitatory and $T_I$ for inhibitory synapses), after which they decay automatically. During this period, the neuron is considered a \emph{ghost} neuron because reactivation does not generate new synaptic activity. Once all outgoing synapses have disappeared, the neuron becomes \emph{fresh} and can again contribute new synaptic influence upon activation. (b) Schematic representation of the network dynamics considered in this work. A small fraction $\eta$ of neurons is activated only at $t=0$, after which the network evolves deterministically through recurrent interactions. Depending on the residual synaptic state, activity may terminate after a single cycle or spontaneously regenerate additional cycles. (c) At the first silent state ($t=t_z^{(1)}$), neuronal activity vanishes while transient synaptic memory persists. The subsequent evolution during the time window $\tau$ is analyzed directly from this synaptic-memory snapshot to quantify the network's latent excitatory recruitment capacity and predict whether activity regeneration will occur. Passing time over $\tau=1,\ldots,T_I$, the gradual decay of residual synaptic activity determines which neurons become eligible for activation. Reactivation of a ghost neuron cannot generate new active synapses. While activation of fresh excitatory neurons has the potential to propagate activity further through the network. }
	\label{fig:ModelSchematic}
\end{figure*}

The network consists of $N$ neurons, divided into $N_E$ excitatory and $N_I$ inhibitory neurons (see Fig.~\ref{fig:ModelSchematic}(a)). The neuronal state is represented by the binary vector
\begin{equation}
	\mathbf{X}(t)=\big( x_1(t),x_2(t),...,x_N(t) \big),
\end{equation}
where
\begin{equation}
	x_i(t)=
	\begin{cases}
		1, & \text{if \textit{i} is active}\\
		0, & \text{if \textit{i} is inactive}
	\end{cases}
\end{equation}
for neuron $i$. In addition to neuronal activity, the model explicitly tracks the remaining lifetime of every synaptic connection. These synaptic states are represented by the set
\begin{equation}
	\mathbf{L}(t)=
	\{L_{ij}(t)\},
\end{equation}
where $L_{ij}(t)$ denotes the remaining active lifetime of the directed synapse from neuron $i$ to $j$. The pair $\big(\mathbf X(t),\mathbf L(t)\big)$ constitutes the complete dynamical state of the network.

\subsection{Synaptic Memory}
The central assumption of the model is that synaptic activation persists for a finite duration. Whenever a neuron becomes active, all of its outgoing synapses are activated with a delay of $d$ and remain active for a prescribed number of time steps $T_E$ and $T_I$ for excitatory and inhibitory synapses, respectively. The lifetime of an active synapse decreases in time until it eventually reaches zero and becomes inactive. Consequently, the network stores information about past neuronal activity within the synaptic state matrix $\mathbf L$. Setting $d=1$ for the sake of simplicity and without loss of generality, this memory mechanism does not require explicit access to previous activity vectors $\mathbf X(t-\tau)$ and $\mathbf L(t-\tau)$ for $\tau > 1$.

\subsection{Neuronal Dynamics} 
\label{sec:NeuronalDynamics}
The total synaptic input received by neuron $i$ is 
\begin{equation}
	I_i(t) =
	W_E
	\sum_{j \in \{ N_E \}}
	A_{ji}
	H \big( L_{ji}(t)\big)
	+
	W_I
	\sum_{j \in \{ N_I \}}
	A_{ji}
	H \big( L_{ji}(t)\big),
\end{equation}
where:
\begin{itemize}
	\item $A_{ij}$ is the adjacency matrix; i.e., $A_{ij}=1$ if pre-synaptic neuron $i$ is connected to post-synaptic neuron $j$, otherwise $A_{ij}=0$,
	\item $W_E$ and  $W_I$ are excitatory and inhibitory synaptic weight, respectively,
	\item $H$ denotes the Heaviside function:
	\begin{equation}
		H(z)=
		\begin{cases}
			1, & z>0\\
			0. & z \le 0
		\end{cases}
	\end{equation}
	
\end{itemize}
Neuronal activation follows
\begin{equation}
	x_i(t)=	
	H \big( I_i(t)-D \big)
	+
	\Big[ 1-H\big(I_i(t)-D \big) \Big] \xi_i(t),
	\label{equ:xi}
\end{equation}
where $D$ is \emph{firing threshold} of the neurons and 
\begin{equation}
	\xi_i(t)
	\sim
	\mathrm{Bernoulli}(\eta).
\end{equation}
In the general formulation, the parameter $\eta$ represents spontaneous neuronal activation originating from external brain regions. Throughout the present study, however, this external drive is applied only at the initial time ($t=0$), where a fraction $\eta$ of neurons is randomly activated to initialize network activity. For all subsequent time steps ($t>0$), the external drive is removed by setting $\eta=0$, such that all later activations arise solely from recurrent synaptic interactions within the network.

When a neuron fires at time $t$, it remains active only during that time step. Its outgoing synapses become active one time step later and remain active for a finite lifetime. Consequently, although the neuron itself might be no longer active, its previous spike continues to affect the network through its still-active outgoing synapses. During this interval, the neuron is said to be in the \emph{ghost state} (see Fig.~\ref{fig:ModelSchematic}(a)). While a neuron remains in the ghost state, any subsequent activation affects only the neuronal state variable $x_i(t)$, however, the associated synaptic counters continue to decay monotonically and are neither reset nor prolonged. 

Once all outgoing synapses have decayed, the neuron no longer leaves any residual influence on the network and returns to the \emph{fresh state}. A fresh neuron can subsequently fire again either by receiving sufficient recurrent synaptic input or, in the general formulation of the model, through spontaneous activation with probability $\eta$. Thus, every neuron continuously alternates between fresh and ghost states throughout the network evolution, depending solely on whether the effect of its previous spike is still present in the network.

It is important to emphasize that the distinction between \emph{ghost} and \emph{fresh} neurons represents an effective dynamical classification rather than additional physiological neuronal states. A \emph{ghost} neuron remains synaptically occupied, meaning that its previously established outgoing synaptic influence is still preserved by finite-lifetime synapses; consequently, additional spikes do not generate new outgoing synaptic effects, even though the neuron's membrane excitability is not assumed to be altered. In contrast, a \emph{fresh} neuron has no remaining active outgoing synaptic trace and is therefore capable of establishing new synaptic influence upon activation. This coarse-grained description characterizes the availability of a neuron's outgoing synaptic influence rather than its membrane excitability, providing an effective representation of the latent synaptic-memory state independently of instantaneous neuronal firing.

\subsection{Synaptic Dynamics}
The evolution of synaptic lifetimes is governed by the following recursive equation: 
\begin{eqnarray}
	L_{ij}(t)=&
	A_{ij}
	\Big[
	\big(L_{ij}(t-1)-1 \big) H\big(L_{ij}(t-1)\big)
	+ \nonumber \\
	&T_{ij}
	x_i(t-1)
	\left(1-H\big(L_{ij}(t-1)\big) \right)
	\Big],
	\label{equ:Lij}
\end{eqnarray}
where the parameter $T_{ij}$ denotes the maximum lifetime associated with the synapse. In the simplest version of the model,
\begin{equation}
	T_{ij}=
	\begin{cases}
		T_E,& i \in \{ N_E\}\\
		T_I,& i \in \{ N_I\}
	\end{cases}
\end{equation}
allowing excitatory and inhibitory synapses to possess different temporal persistence. The factor $A_{ij}$ in Eq.~(\ref{equ:Lij}) guarantees that counters of lifetime change only for links that actually exist in the adjacency matrix. The first term describes the evolution of links whose activity has already started, while the second term activates previously inactive links whenever the corresponding presynaptic neuron was active one time step earlier.

\subsection{State-Space Representation and Macroscopic Observables}
The microscopic state of the network at time $t$ is completely specified by the neuronal state vector
$\mathbf{X}(t)$ and the synaptic memory matrix
$\mathbf{L}(t)$. Accordingly, the complete state of the system can be written as
\begin{equation}
	\mathbf{S}(t)=\left(\mathbf{X}(t),\mathbf{L}(t)\right).
\end{equation}
In the general form of the model with randomly activation of neurons at each time step, the system evolves according to 
\begin{equation}
	\mathbf{S}(t+1)=F\!\left(\mathbf{S}(t) , \mathbf{\Xi} (t+1) \right),
\end{equation}
where $\mathbf{\Xi} (t) = \left( \xi_1(t), \ldots, \xi_N (t) \right)$. However, for the present study where spontaneous activation is applied only at the initial time $t=0$ (see Fig.~\ref{fig:ModelSchematic}(b)), we have 
\begin{equation}
	\mathbf{S}(t+1)=F\!\left(\mathbf{S}(t)\right),
\end{equation}
Here $F$ denotes the deterministic update operator since both neuronal activation and synaptic evolution are fully determined by the update rules introduced in Eq.~(\ref{equ:xi}) and Eq.~(\ref{equ:Lij}). Throughout the present work, at $t>0$, the subsequent evolution is entirely governed by the internal dynamics of the network.

The macroscopic observables used in this study are the density of active neurons,
\begin{equation}
	\rho(t)=
	\frac{1}{N}
	\sum_{i=1}^{N}
	x_i(t),
\end{equation}
and the density of active excitatory and inhibitory synapses,
\begin{equation}
	\phi_\alpha(t)=
	\frac{1}{l}
	\sum_{i \in \{ N_\alpha \}}
	\sum_{j=1}^{N}
	H\big(L_{ij}(t)\big),
\end{equation}
with $\alpha \in \{E,I\}$ and $l=\sum_{i=1}^{N} \sum_{j=1}^{N} A_{ij}$ as the total number of directed synaptic connections in the network. Consequently, the density of total active synapses is obtained as $\phi(t) = \phi_E(t) + \phi_I(t)$.  

An important consequence of this formulation is that the future evolution of the system depends on both neuronal activity and synaptic memory. Two realizations may exhibit identical neuronal activity at a given time, i.e.,
$\mathbf{X}_1(t)=\mathbf{X}_2(t)$,
while possessing different synaptic memory matrices,
$\mathbf{L}_1(t)\neq\mathbf{L}_2(t)$.
Such realizations generally evolve toward different future trajectories, demonstrating that neuronal activity alone is insufficient to determine the subsequent dynamics. In other words, the synaptic memory matrix provides additional hidden state variables that govern the evolution of the system. Consequently, the synaptic memory matrix acts as an internal memory of the system, allowing the network to distinguish between microscopic states that would otherwise appear identical if only neuronal activity were considered.

\section{Results} 
\label{sec:results}
\begin{figure*}[t!]
	\centering    \includegraphics[width=0.9\linewidth]{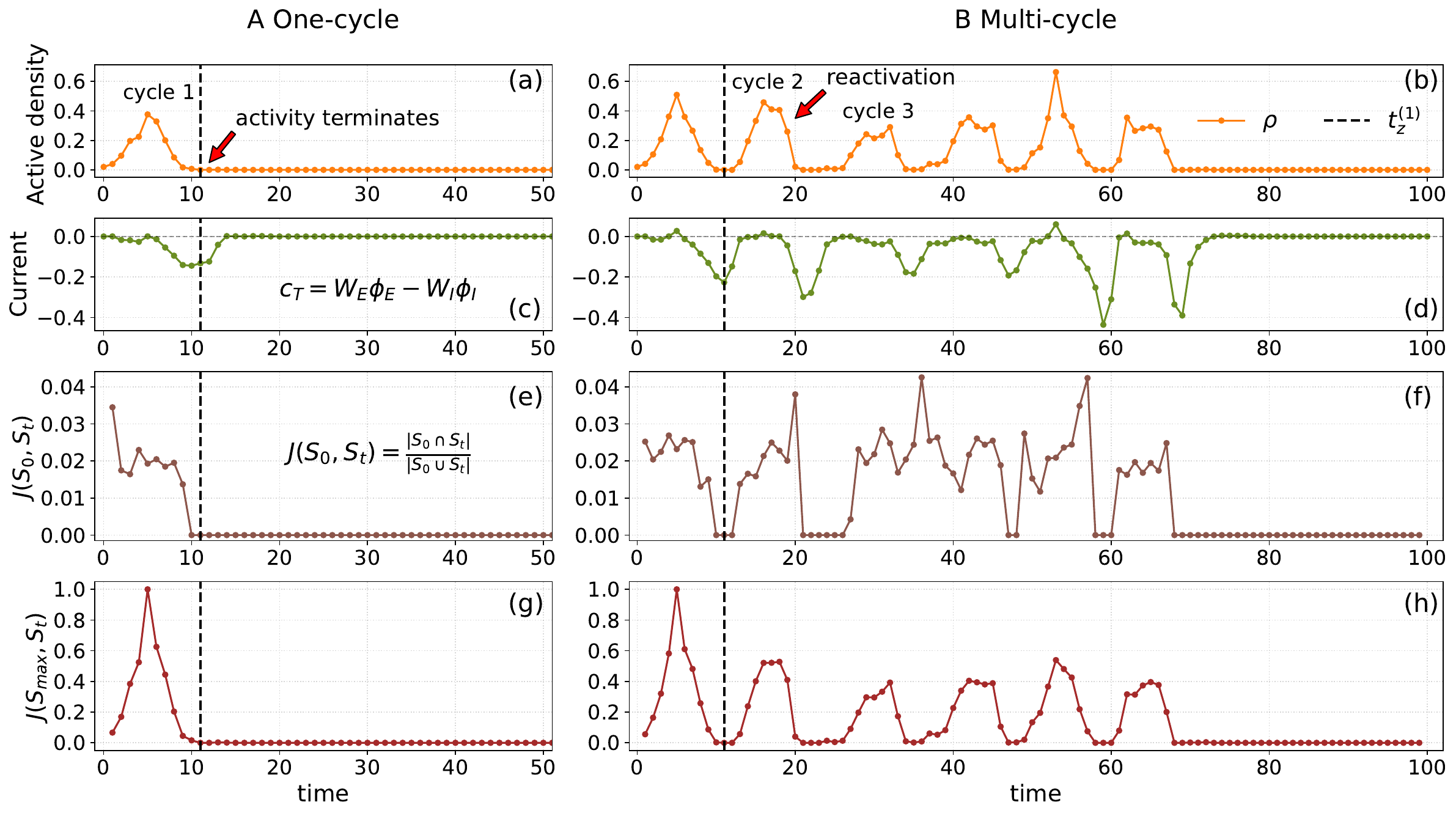}
	\caption{\textbf{Representative examples of the one-cycle and multi-cycle dynamical regimes together with several macroscopic quantities that fail to distinguish between them.}
		(a,b) Temporal evolution of the active neuronal density $\rho(t)$. In the one-cycle regime, activity terminates after the first silent interval at $t=t_z^{(1)}$, whereas in the multi-cycle regime activity spontaneously regenerates without additional external input.
		(c,d) Evolution of the global synaptic current, $c_T=W_E\phi_E-W_I\phi_I$, showing that this macroscopic quantity alone does not predict activity regeneration.
		(e,f) Jaccard similarity between the active neuronal assembly and the initially stimulated assembly $S_0$. In both regimes, the overlap rapidly decays, indicating that later activity is carried by different neuronal assemblies.
		(g,h) Jaccard similarity with the largest assembly formed during the first cycle ($S_5$). Unlike $J(S_0,S_t)$, this overlap remains substantially higher, suggesting that subsequent activity is organized around dynamically formed assemblies, consistent with the concept of dynamic coding.
		Overall, these macroscopic measures fail to distinguish between one-cycle and multi-cycle dynamics, motivating the search for a more informative predictor based on the residual synaptic state.}
	\label{fig:OneMulti}
\end{figure*}

The network considered in this work is intentionally simple from a structural point of view. Excitatory and inhibitory neurons are connected through a random directed graph with $N=2000$, $N_E = 4N_I$ and homogeneous in-degree and out-degree distributions with a mean of $\langle k_{in} \rangle = \langle k_{out} \rangle = 200$. The activation threshold of neurons is set as $D=5$. Furthermore, all excitatory synapses share the same positive weight $W_E=1$ and maximum lifetime $T_E=5$, while all inhibitory synapses share the negative weight $W_I=-4$ and longer lifetime $T_I=7$ \cite{khanjanianpak2024emergence}. Consequently, the network contains neither predefined cell assemblies nor structural heterogeneity that could trivially explain different dynamical behaviors. The implementation used in this work is publicly available (see Materials Availability in Sec.~\ref{sec:material}).

Although the underlying network is structurally homogeneous, repeated realizations reveal two qualitatively distinct dynamical outcomes. Starting from different realizations of the same initial activation density, the network may either generate only a single activation cycle or spontaneously produce one or more additional activation cycles after the first silent state. 

Notably, the emergence of these two responses cannot be attributed to predefined structural motifs, heterogeneous degree distribution and/or synaptic strengths, or specialized neuronal populations. It seems that the only source of variability between different realizations is the random selection of the initially activated neurons. Understanding the origin of these two outcomes constitutes the central question of the present work.

\subsection{Initial Observations and the Search for a Predictor}
To investigate how transient neuronal activity can give rise to self-sustained reactivation, we initialized the network by activating a random fraction $\eta =0.02$ of neurons only at $t=0$. After this initial perturbation, the external drive was completely removed ($\eta=0$), and the subsequent dynamics evolved solely through recurrent synaptic interactions and the finite lifetime of active synapses.

The two qualitatively distinct dynamical behaviors are demonstrated in  Fig.~\ref{fig:OneMulti} as representative examples.  In the first case, Fig.~\ref{fig:OneMulti}(a), neuronal activity gradually decreases and eventually reaches zero at a finite time, denoted by $t_z^{(1)}$. Once the network becomes silent, no further activity emerges, and the evolution terminates after a single activation cycle. We refer to this behavior as the \emph{one-cycle} state.

In the second case, Fig.~\ref{fig:OneMulti}(b), the neuronal activity also reaches zero at $t_z^{(1)}$, but after a silent interval a new activation cycle spontaneously emerges without any external stimulation. We call this type of behavior as \emph{multi-cycle} state. Depending on the realization, this spontaneous reactivation may occur only once or may repeat several times, producing multiple activation cycles. Throughout this work, however, our primary objective is to understand the mechanism responsible for the emergence of the \emph{next} activation cycle immediately following the first silent state. Once this transition is understood, the same mechanism naturally applies to subsequent cycles as well.

The existence of these two behaviors naturally raises the following question: What determines whether the network remains permanently silent after the first extinction of activity or spontaneously generates the next activation cycle? To answer this question, we first examined several intuitive macroscopic descriptors of the network dynamics. Surprisingly, none of them was able to reliably discriminate between the one-cycle and multi-cycle regimes.

Figures~\ref{fig:OneMulti}(c,d) compare the evolution of the total synaptic current,
$c_T=W_E\phi_E - W_I\phi_I$,
for representative one-cycle and multi-cycle realizations. Although the temporal profiles differ, no characteristic behavior capable of consistently separating the two dynamical regimes could be identified.

We also examined the temporal evolution of active neuronal assemblies using the Jaccard similarity index \cite{jaccard1901etude}, defined as:
\begin{equation}
	J(S_{A},S_{B})= \frac{|S_{A}\cap S_{B}|}{|S_{A}\cup S_{B}|},
	\label{equ:jaccardindex}
\end{equation}
where symbols $\cap$ and $\cup$ denote the intersection and the union of sets $S_A$ and $S_B$ as the active assemblies at time $t_A$ and $t_B$, respectively. As depicted in Fig.~\ref{fig:OneMulti}(e,f), the similarity between the active assembly at time $t$ and the initially stimulated assembly $S_0$, i.e., $J(S_0,S_t)$, rapidly decays to values close to zero in both dynamical regimes. Thus, even when spontaneous reactivation occurs, it is generally not produced by reactivation of the original neuronal population.

In contrast, the similarity with the assembly formed near the peak of the first activation cycle ($S_{\max}$; corresponding to $S_5$ in the representative examples shown here) remains substantially larger (Fig.~\ref{fig:OneMulti}(g,h)). This observation indicates that subsequent activity is preferentially organized around neuronal assemblies that emerge dynamically during network evolution rather than around the initially stimulated assembly. Such continuous turnover of active neuronal populations is consistent with the concept of dynamic coding \cite{stroud2024computational}, in which information is maintained by evolving neuronal assemblies instead of persistent activity within a fixed population.

Although these observations provide useful insight into the collective organization of network activity, neither the global synaptic current nor the similarity between neuronal assemblies provides a reliable predictor for the emergence of the next activation cycle. This motivated us to search for a quantity more directly related to the microscopic mechanism responsible for spontaneous reactivation.

\subsection{The Role of Fresh Excitatory Neurons}

\begin{figure*}[t!]
	\centering    \includegraphics[width=0.88\linewidth]{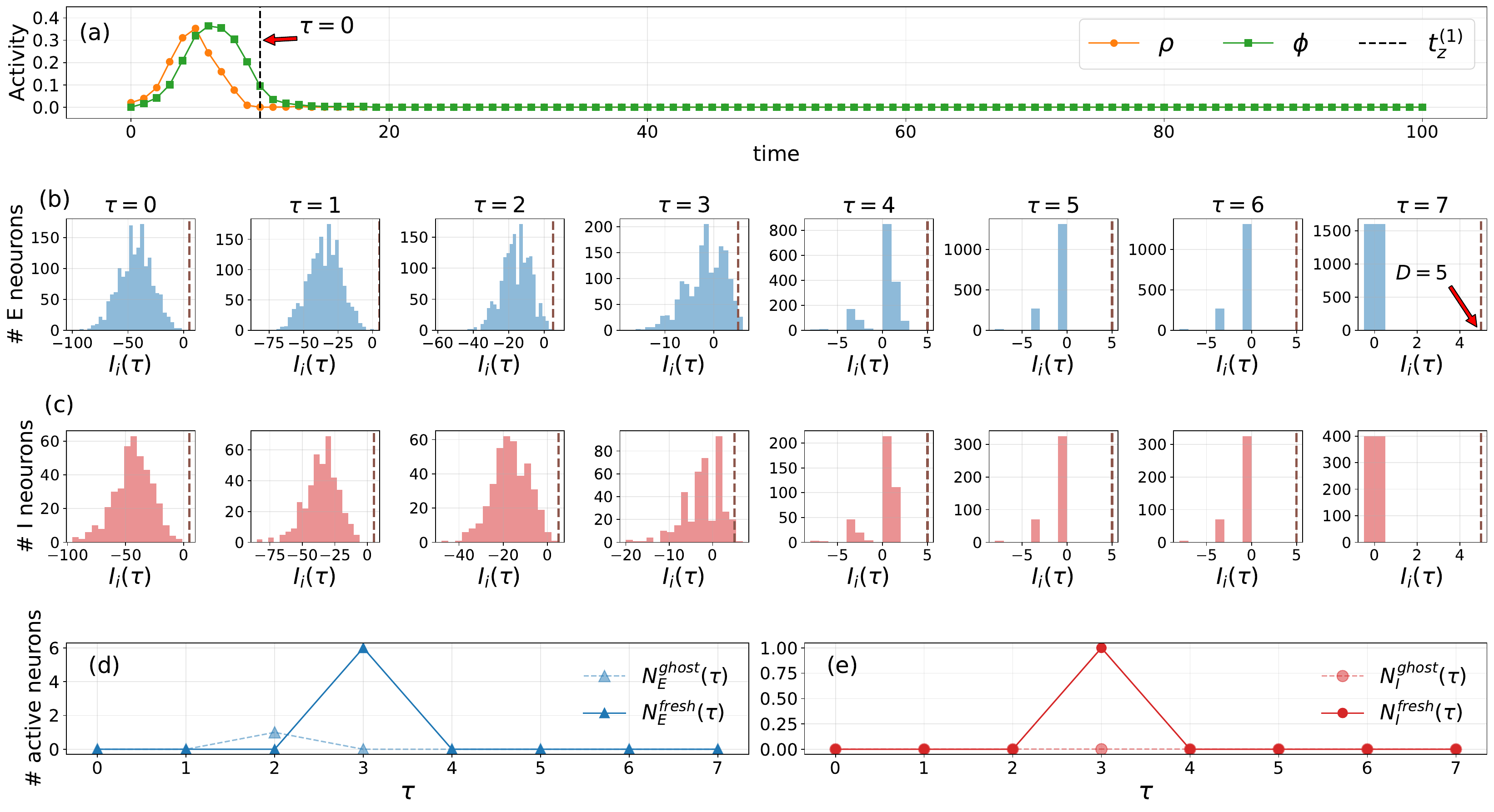}
	\caption{ \textbf{One-cycle example.} 	(a) Active density of neurons $\rho$ and synapses $\phi$ in time. Although neuronal activity vanishes at the first silent state $t_z^{(1)}$ (dashed vertical line), a substantial fraction of synapses remains active, motivating the snapshot analysis. (b,c) Evolution of the reconstructed excitatory and inhibitory input-current distributions obtained solely from the synaptic-memory snapshot at $t_z^{(1)}$, without continuing the simulation. Dashed vertical line indicates activation threshold of $D=5$. Numbers of reactivated ghost and fresh (d) excitatory (e) inhibitory neurons during $\tau=t-t_z^{(1)}$. The gradual decay of residual synaptic activity determines which neurons become eligible for reactivation, with fresh neurons largely dominating over ghost neurons, indicating that the network's capacity for generating new activity is primarily carried by fresh neurons. }
	\label{fig:HistogramOne}
\end{figure*}

\begin{figure*}[t!]
	\centering    \includegraphics[width=0.88\linewidth]{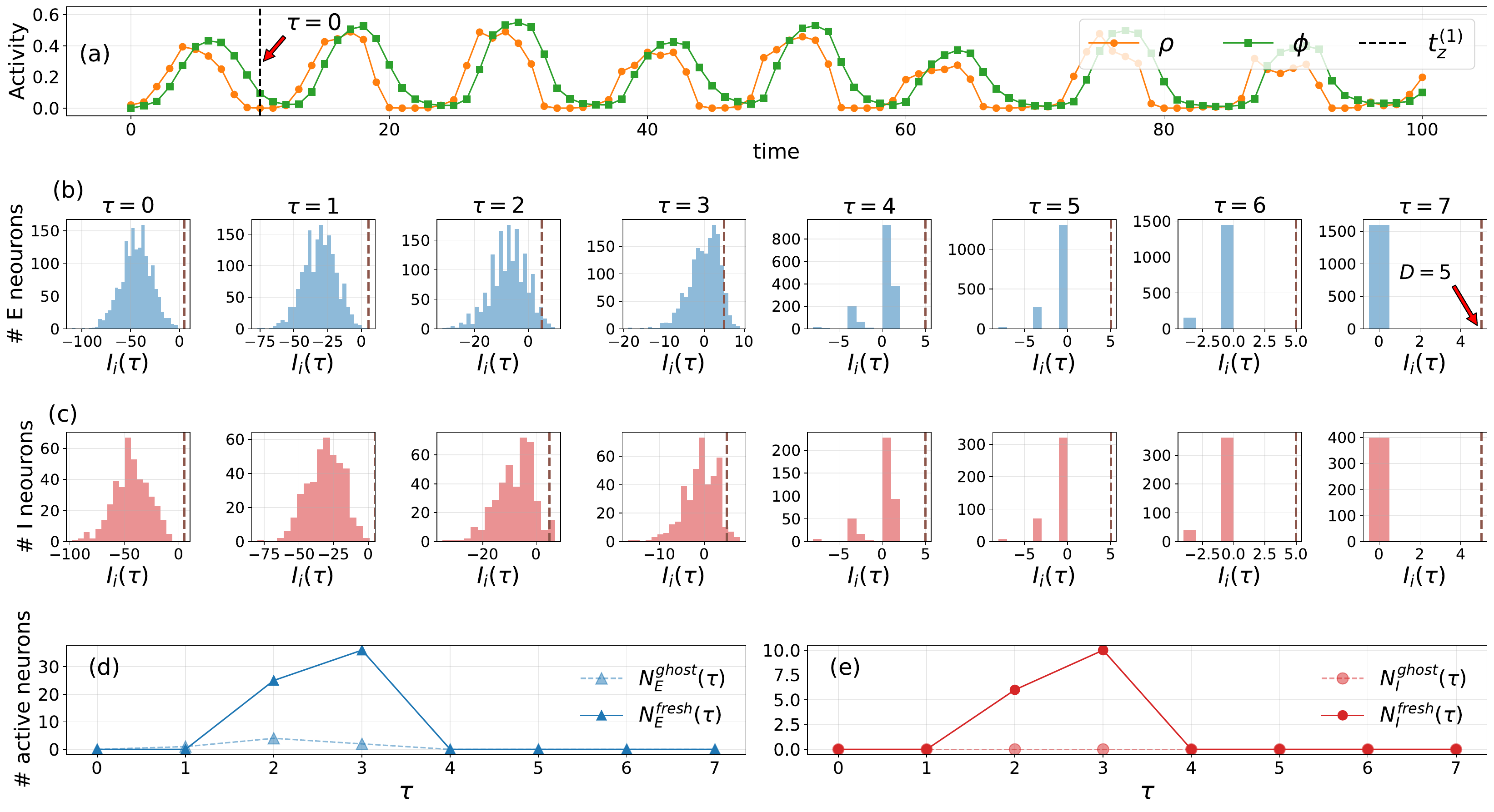}
	\caption{\textbf{Multi-cycle example.} (a) Active density of neurons $\rho$ and synapses $\phi$ in time. (b,c) Evolution of the reconstructed excitatory and inhibitory input-current distributions obtained solely from the synaptic-memory snapshot at $t_z^{(1)}$. Dashed vertical line indicates activation threshold of $D=5$. Numbers of reactivated ghost and fresh (d) excitatory (e) inhibitory neurons during $\tau=t-t_z^{(1)}$. Although the current distributions span ranges comparable to those in the one-cycle example (Fig.~\ref{fig:HistogramOne}), substantially more neurons exceed the activation threshold, suggesting a larger latent capacity for generating the next activation cycle.}
	\label{fig:HistogramMulti}
\end{figure*}

The negative results presented in the previous subsection suggest that the transition from the one-cycle to the multi-cycle regime cannot be inferred from global macroscopic observables alone. We therefore sought a quantity more directly related to the microscopic mechanism responsible for initiating the next activation cycle.

A key observation is that the network is not completely inactive when the neuronal activity first reaches zero (see Fig.~\ref{fig:ModelSchematic}(c)). Let $t_z^{(1)}$ denote the first time at which
\begin{equation}
	\rho(t_z^{(1)})=0,
\end{equation}
implying that
\begin{equation}
	\mathbf{X}(t_z^{(1)})=0.
\end{equation}
Although no neuron is active at this instant, a considerable fraction of synapses remain active, as shown in Fig.~\ref{fig:HistogramOne}(a) and Fig.~\ref{fig:HistogramMulti}(a), i.e., 
\begin{equation}
	\phi(t_z^{(1)}) \neq 0.
\end{equation}
Consequently, the synaptic memory matrix
$\mathbf{L}(t_z^{(1)})$
still stores information about the previous activity and completely determines how the residual synaptic currents will evolve until all active synapses decay.

Rather than continuing the original simulation, we therefore regard
$\mathbf{L}(t_z^{(1)})$
as a \emph{snapshot} of the network and use it as the initial condition for a new analysis. Since the longest synaptic lifetime is that of inhibitory synapses ($T_I=7$), every synapse present in this snapshot must disappear within at most $T_I$ subsequent time steps. Hence it is sufficient to analyze the interval
\[
\tau=0,\ldots,T_I,
\]
where $\tau=t-t_z^{(1)}$.

Using only the information stored in
$\mathbf{L}(t_z^{(1)})$,
the synaptic input received by neuron $i$ after $\tau$ time steps is
\begin{eqnarray}
	I_i(t_z^{(1)}+\tau)
	& = 
	W_E
	\sum_{j\in \{N_E\} } A_{ji}
	H\!\left(L_{ji}(t_z^{(1)})-\tau\right) - \nonumber \\
	& W_I
	\sum_{j\in \{ N_I\}} A_{ji} 
	H\!\left(L_{ji}(t_z^{(1)})-\tau\right),
	\label{eq:SnapshotCurrent}
\end{eqnarray}
where $H(L_{ji}(t_z^{(1)})-\tau)$
indicates whether the synapse from neuron $j$ to neuron $i$ is still active after $\tau$ time steps. Thus, Eq.~(\ref{eq:SnapshotCurrent}) describes the gradual disappearance of the residual synaptic currents stored in the snapshot, without introducing any new synaptic activity.

According to the neuronal dynamics introduced in Sec.~\ref{sec:NeuronalDynamics}, neuron $i$ has the potential to become active at delay $\tau$ whenever
\begin{equation}
	I_i(t_z^{(1)}+\tau)>D.
\end{equation}
However, not every such neuron can contribute to the formation of the next activation cycle.

At this point it is useful to distinguish between neurons in the ghost and fresh states. Ghost neurons are analogous to recalling a verse of a poem that is already being repeated in one's mind: repeating it again does not create any additional memory trace. Likewise, reactivation of a ghost neuron cannot generate new active synapses because its previous outgoing synapses are still active. In contrast, activation of a fresh neuron might create a completely new set of active synapses and therefore has the potential to propagate activity further through the network (see Fig.~\ref{fig:ModelSchematic}(c)).

A neuron remains in the ghost state whenever at least one of its outgoing synapses is still active. Since the synaptic lifetime $L_{ij}$ is strictly positive only while the synapse is active, the sum of all outgoing counters of neuron $i$ is positive if and only if at least one outgoing synapse remains active. Therefore, neuron $i$ is in the ghost state at delay $\tau$ if
\begin{equation}
	H \Big( \sum_{j=1}^{N} A_{ij} \big( L_{ij}(t_z^{(1)})-\tau \big) \Big) =1.	
	\label{eq:GhostCondition}
\end{equation}
Conversely, a neuron is fresh only when all of its outgoing synapses have completely decayed.

Using Eqs.~(\ref{eq:SnapshotCurrent}) and (\ref{eq:GhostCondition}), the number of fresh excitatory neurons that are capable of becoming active at delay $\tau$ is
\begin{align}
	N_E^{\mathrm{fresh}}(\tau)
	&=
	\sum_{i\in \{N_E \}}
	\Big[
	H\Big(I_i(t_z^{(1)}+\tau)-D\Big) \times
	\nonumber\\
	&\qquad
	\left(
	1-
	H\Big(
	\sum_{j=1}^{N} A_{ij}
	\big(L_{ij}(t_z^{(1)})-\tau\big)
	\Big)
	\right)
	\Big].
	\label{eq:FreshExcitatory}
\end{align}
and an analogous expression is used for inhibitory neurons when $i \in \{ N_I\}$.

The evolution of the input-current distributions obtained from Eq.~(\ref{eq:SnapshotCurrent}) is shown in Fig.~\ref{fig:HistogramOne}(b,c) and Fig.~\ref{fig:HistogramMulti}(b,c) for representative one-cycle and multi-cycle realizations. As inhibitory synapses gradually decay, an increasing number of neurons might exceed the firing threshold. Figures~\ref{fig:HistogramOne}(d,e) and \ref{fig:HistogramMulti}(d,e) further demonstrate that the overwhelming majority of these neurons are in the fresh state, whereas only a small fraction belongs to the ghost state. Therefore, the network's ability to initiate the next activation cycle is determined primarily by the activation of fresh neurons rather than by repeated activation of ghost neurons.

Since only fresh excitatory neurons are capable of establishing new excitatory synapses and propagating activity further through the network, this quantity turns out to be the key predictor of whether the network will remain in the one-cycle state or spontaneously generate the next activation cycle.

\begin{figure*}[t!]
	\centering    \includegraphics[width=0.95\linewidth]{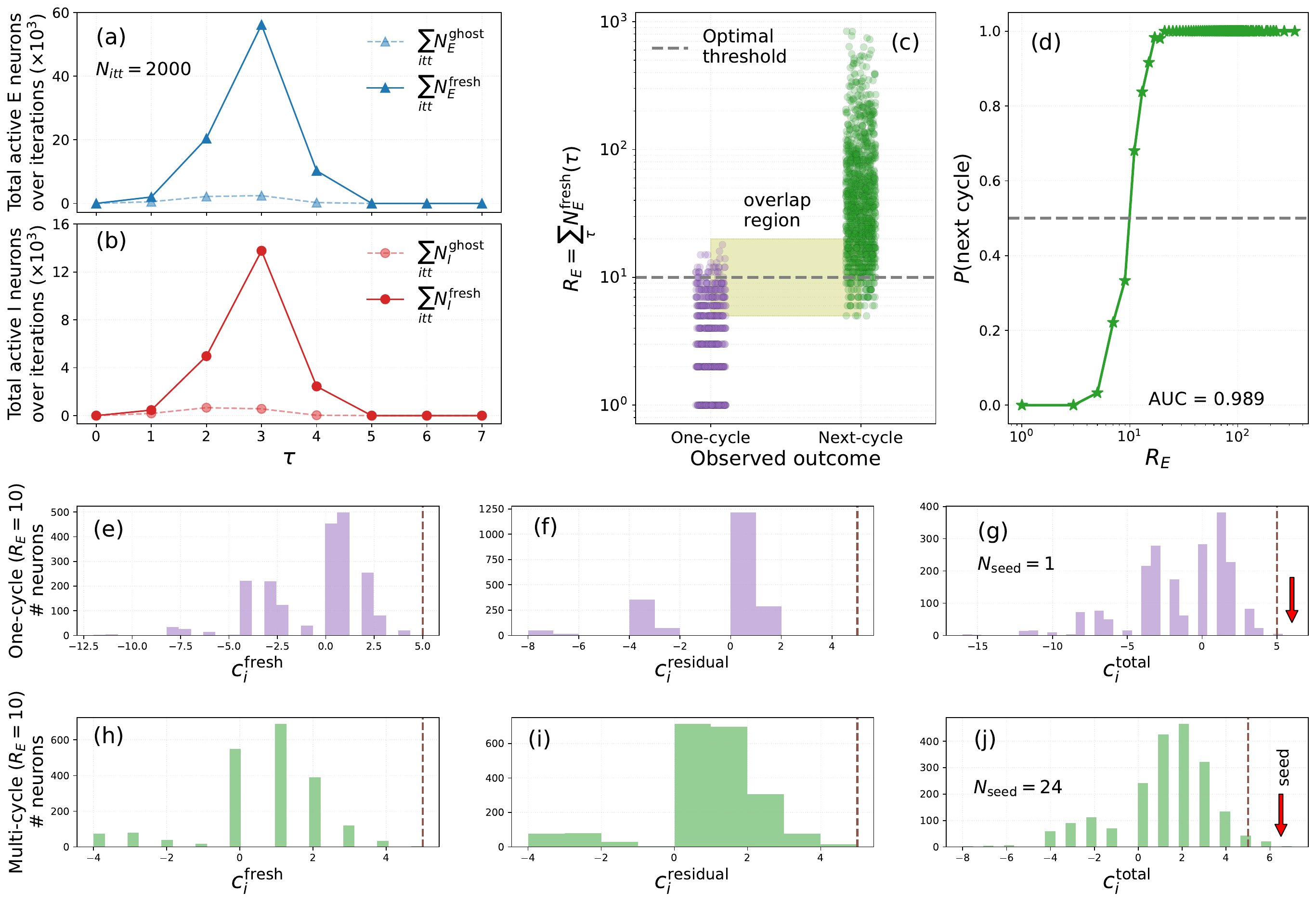}
	\caption{\textbf{Prediction of multi-cycle emergence from the synaptic-memory snapshot.} (a,b) Temporal evolution of the numbers of fresh excitatory and inhibitory neurons that become eligible for activation during $\tau=t-t_z^{(1)}$, aggregated over 2000 independent realizations. Most activation events occur within the first few time steps after the first silent state, indicating that the Latent Excitatory Recruitment (LER) of the network is established over a short time window. (c) Classification of all realizations according to the cumulative number of fresh excitatory neurons, 		$R_E=\sum_{\tau=0}^{T_I}N_E^{\mathrm{fresh}}(\tau)$. Each point corresponds to one realization and is labeled according to whether the original network subsequently exhibited one-cycle or multi-cycle dynamics. Three distinct regions emerge: a one-cycle region for small $R_E$, a multi-cycle region for large $R_E$, and a narrow overlap region where both outcomes are possible. (d) Conditional probability $P(\mathrm{multi}\,|\,R_E)$, demonstrating that $R_E$ serves as an excellent predictor of multi-cycle emergence. (e--j) Two representative realizations from the overlap region with identical $R_E$ but different dynamical outcomes. The distributions of the synaptic current generated by freshly recruited neurons ($c_i^{fresh}$), the residual synaptic-memory current ($c_i^{residual}$), and their sum ($c_i^{total}$) are compared for one-cycle (e--g) and multi-cycle (h--j) realizations. Despite the identical latent excitatory reactivation capacity, the different current distributions generate vastly different numbers of activation seeds ($N_{seed}=1$ versus $24$), explaining why only the latter successfully regenerates a new collective activity cycle.}
	\label{fig:PmultiOverlap}
\end{figure*}

\subsection{Probabilistic Prediction of Multi-Cycle Emergence}
The snapshot-based analysis described in the previous subsection was repeated for all 2000 independent realizations generated from different random initial activation patterns. For each realization, the synaptic-memory snapshot at the first silent state, $\mathbf{L}(t_z^{(1)})$, was extracted and analyzed independently of the subsequent network dynamics. Consequently, all quantities presented in this section are obtained solely from the information contained in the snapshot rather than by continuing the original simulation.

Figure~\ref{fig:PmultiOverlap}(a,b) summarizes the temporal evolution of the numbers of fresh excitatory and inhibitory neurons that become eligible for activation during the interval $\tau=0,\ldots,T_I$. In most realizations, the majority of these activation events occur within the first few time steps after the silent state, with a pronounced maximum around $\tau\approx3$. This characteristic timescale reflects the progressive decay of the residual inhibitory synapses stored in the snapshot, which gradually releases neurons from inhibition. After approximately $\tau=T_I$, virtually no additional neurons become eligible for activation because all synaptic memory retained in $\mathbf{L}(t_z^{(1)})$ has vanished. 

This naturally motivates introducing the cumulative quantity
\begin{equation}
	R_E
	=
	\sum_{\tau=0}^{T_I}
	N_E^{\mathrm{fresh}}(\tau),
\end{equation}
which measures the \emph{latent excitatory recruitment} (LER) encoded in the synaptic-memory snapshot. In other words, $R_E$ quantifies how many fresh excitatory neurons can potentially be recruited solely by the residual synaptic memory stored in $\mathbf{L}(t_z^{(1)})$.

Figure~\ref{fig:PmultiOverlap}(c) presents the outcome of all realizations as a function of $R_E$. Each point corresponds to one realization and is classified according to whether the original network subsequently exhibited only a single activation cycle (one-cycle) or generated one or more additional activation cycles (multi-cycle). Remarkably, despite the large variability of the initial activation patterns, the realizations organize into two well-separated regions.

For small values of $R_E$, only one-cycle behavior is observed, indicating that the LER stored in the snapshot is insufficient to initiate another activation cycle. Conversely, above a critical range of $R_E$, every realization develops at least one additional activation cycle. Between these two regimes, however, an overlap region exists in which both outcomes become possible.

To quantify this transition, Fig.~\ref{fig:PmultiOverlap}(d) shows the conditional probability
\[
P(\mathrm{multi}\mid R_E),
\]
computed from all realizations, such that for each value of $R_E$ the probability of observing multi-cycle state, $P(\mathrm{multi}|R_E)$, equals to the number of multi-cycle observation over total number of realizations. This probability exhibits a monotonic sigmoidal-like increase from nearly zero to unity, demonstrating that the LER provides an excellent predictor of whether the network escapes the first silent state and generates another activation cycle.

The overlap region deserves particular attention because realizations with identical values of $R_E$ may still evolve toward different dynamical outcomes. Two representative examples are presented in Figs.~\ref{fig:PmultiOverlap}(e--g) and \ref{fig:PmultiOverlap}(h--j), where the two realizations possess the same LER but eventually produce one-cycle and multi-cycle dynamics, respectively (also see Figs~\ref{fig:HistogramOvrelapOne} and \ref{fig:HistogramOverlapMulti}). This observation indicates that although $R_E$ captures the principal control parameter governing the transition, finer details of how the synaptic drive is partitioned between newly recruited neurons and the residual synaptic-memory state may still influence the final outcome when the system operates close to the transition boundary.

To further illustrate this point, we decomposed the total synaptic input received by each neuron during the overlap region into two distinct components. The first component, shown in Figs.~\ref{fig:PmultiOverlap}(e,h) and denoted by $c_i^{fresh}$, represents the synaptic current generated exclusively by the newly recruited (fresh) neurons that become active at the corresponding value of $\tau$, which equals $\tau=3$ for the representative pair as can be seen in Figs.~\ref{fig:HistogramOvrelapOne} and \ref{fig:HistogramOverlapMulti}. The second component, $c_i^{residual}$, shown in Figs.~\ref{fig:PmultiOverlap}(f,i), corresponds to the current carried by synapses that remain active from the original activity cycle. Their sum, $c_i^{total} = c_i^{fresh} + c_i^{residual}$, shown in Figs.~\ref{fig:PmultiOverlap}(g,j), represents the instantaneous synaptic drive experienced by neuron $i$. Although the two realizations possess the same value of $R_E$, the distributions of these current components differ noticeably. Consequently, the number of neurons receiving suprathreshold synaptic input and capable of acting as seeds for the next activation cycle differs dramatically, with $N_{seed} =1$ for the one-cycle realization and $N_{seed=}24$ for the multi-cycle realization. These observations indicate that, close to the transition boundary, not only the amount of latent excitatory recruitment but also the spatial organization of the recruited neurons and their interaction with the residual synaptic memory determine whether a sufficient number of activation seeds can be established to initiate another collective activity cycle.

Outside this narrow overlap region, however, $R_E$ alone provides an almost deterministic prediction of the network response. The corresponding receiver operating characteristic yields an area under the curve of  $\mathrm{AUC} = 0.989$, confirming the remarkable predictive power of this simple snapshot-derived quantity which suggests $R_E$ might to be considered as an \emph{order parameter} instead of a simple predictor.

\begin{figure*}[t!]
	\centering    \includegraphics[width=1\linewidth]{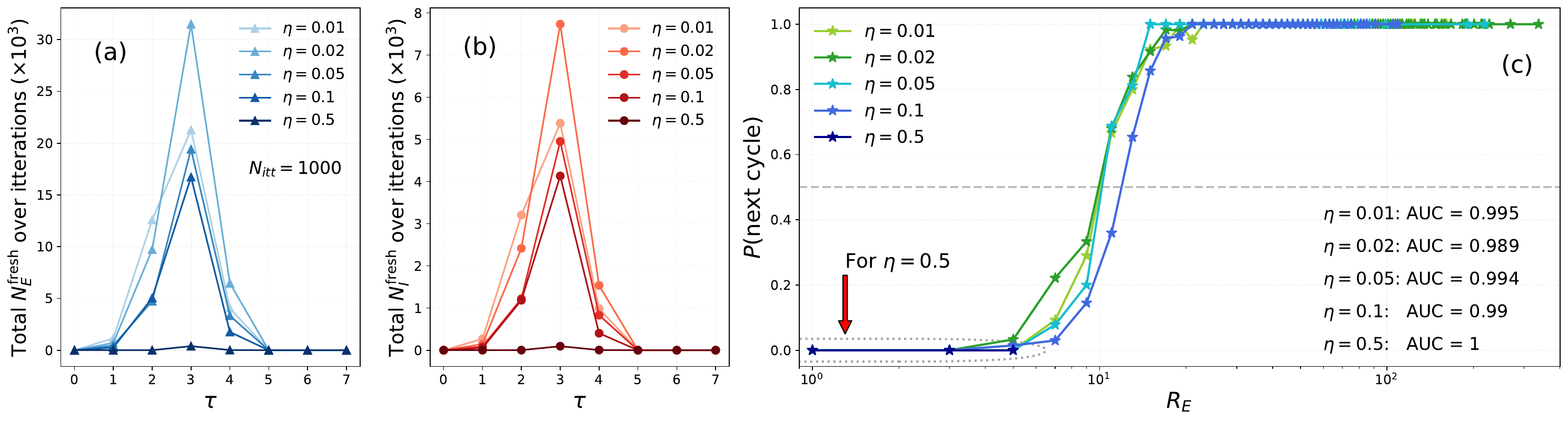}
	\caption{\textbf{Robustness of the proposed predictor against variations of the initial activation density.} 	(a,b) Total numbers of fresh excitatory and inhibitory neurons obtained from the snapshot analysis, aggregated over 1000 independent realizations for each value of the initial activation density $\eta$. Increasing the initial activation density does not enhance the network's ability to sustain activity; instead, it systematically reduces the number of fresh neurons available for recruitment after the first silent state, indicating that stronger initial activation leaves a larger fraction of neurons trapped in the ghost state. (c) Conditional probability of observing at least one additional activation cycle as a function of the cumulative number of fresh excitatory neurons, 		$R_E=\sum_{\tau=0}^{T_I}N_E^{\mathrm{fresh}}(\tau)$, for different values of $\eta$. The curves for $\eta=0.01$--$0.1$ nearly collapse onto one another and yield consistently high AUC values, demonstrating that the predictive power of $R_E$ is essentially independent of the initial activation density. In contrast, for the extreme case $\eta=0.5$ (red arrow), multi-cycle behavior is completely absent because the snapshot generates too few fresh excitatory neurons to exceed the critical reactivation capacity required for initiating another activation cycle. 	}
	\label{fig:EtaEffect}
\end{figure*}

\subsection{Robustness of the Predictor with Respect to the Initial Activation Density}

To examine whether the proposed predictor depends on the initial level of network activation, the entire snapshot-based analysis was repeated for several values of the initial activation density,
$\eta=0.01$, $0.02$, $0.05$, $0.1$, and $0.5$. For each value of $\eta$, 1000 independent realizations were generated, and the corresponding synaptic-memory snapshots at the first silent state were analyzed following exactly the same procedure described in the previous sections.

Figures~\ref{fig:EtaEffect}(a,b) summarize the total numbers of fresh excitatory and inhibitory neurons generated during the snapshot analysis. As the initial activation density increases, the total numbers of fresh neurons decrease systematically for both neuronal populations. This trend is initially counterintuitive, since one might expect that activating more neurons at the beginning would facilitate subsequent reactivation. Instead, the opposite occurs. A larger initial perturbation simultaneously activates a larger fraction of synapses, causing many neurons to remain in the ghost state for a longer period. As a result, fewer neurons become available for fresh recruitment during the subsequent decay of the synaptic memory, reducing the network's LER.

Despite this quantitative change, the predictive role of $R_E$ remains remarkably robust. Figure~\ref{fig:EtaEffect}(c) shows the conditional probability
$P(\mathrm{multi}\mid R_E)$
for all investigated values of $\eta$. For the physiologically relevant range
$\eta=0.01$--$0.1$, the probability curves almost completely collapse onto one another despite the different initial activation levels. Likewise, the corresponding ROC analysis yields AUC values between approximately $0.99$ and $1.00$, demonstrating that the predictive power of $R_E$ is essentially independent of the initial activation density. This collapse indicates that the probability of generating an additional activation cycle is governed primarily by the LER encoded in the snapshot rather than by the magnitude of the initial perturbation itself.

A qualitatively different behavior appears only for the extreme case of
$\eta=0.5$, highlighted by the red arrow in Fig.~\ref{fig:EtaEffect}(c). In this regime, multi-cycle behavior is completely absent and every realization terminates after the first activation cycle, resulting in an AUC equal to unity. Examination of the corresponding snapshots reveals that only a very small number of fresh excitatory neurons can be recruited, placing all realizations entirely within the one-cycle region. This observation provides strong support for the proposed mechanism: when the initial perturbation is excessively large, most neurons spend a prolonged period in the ghost state because of the widespread synaptic activation created during the first cycle. Consequently, the network loses its ability to recruit a sufficient number of fresh excitatory neurons to initiate another activation cycle.

Taken together, these results demonstrate that the predictor introduced in this work is not tied to a particular choice of the initial activation density. Instead, the decisive quantity is the LER stored in the synaptic-memory snapshot, regardless of how that snapshot was generated. These results suggest that the network does not remember because many neurons were initially activated. Rather, it remembers only when the synaptic memory retained after the first silent state is capable of recruiting a sufficiently large population of fresh excitatory neurons.

\section{Discussion} 
\label{sec:discussion}
The present study addresses a fundamental question concerning neuronal networks with transient synaptic memory: why do some activity episodes terminate after a single activation cycle, whereas others spontaneously regenerate one or more additional cycles despite evolving under the same network architecture and identical dynamical rules? At first sight, the coexistence of these two behaviors appears highly counterintuitive. Since the underlying connectivity, synaptic parameters, and neuronal dynamics remain unchanged, one might expect that the network should respond similarly whenever comparable levels of activity are generated. Our results demonstrate, however, that this intuition is incomplete. The decisive information is not contained in the instantaneous neuronal activity itself but rather in the transient synaptic state that survives after neuronal firing has completely ceased. This observation naturally extends the idea of activity-silent memory, in which transient synaptic states retain information after neuronal firing has subsided \cite{mongillo2008synaptic,stokes2015activity}.

A central contribution of this work is the introduction of a snapshot-based description of network dynamics. Instead of following the full temporal evolution after the first silent state, we analyzed only the residual synaptic-memory configuration existing at that instant. This perspective revealed that the first silent state should not be regarded as the termination of network dynamics. Although all neurons are inactive, a substantial population of synapses remains active because of their finite lifetimes, preserving a transient record of the preceding activity. Consequently, the future evolution of the network is already partially encoded within this residual synaptic state.

Another important aspect of the present results is that the coexistence of one-cycle and multi-cycle dynamics emerges in a structurally homogeneous network. No heterogeneity in connectivity, synaptic strength, or neuronal parameters is required. Instead, the diversity of network trajectories arises solely from the transient synaptic-memory state inherited from previous activity. This finding highlights that transient synaptic memory itself can act as a source of functional diversity, even in otherwise homogeneous neuronal circuits. While previous studies have emphasized transient synaptic memory primarily as a substrate for short-term information storage \cite{mongillo2008synaptic, stokes2015activity}, our results further suggest that it can also generate distinct dynamical trajectories even in structurally homogeneous networks.

Systematic examination of numerous candidate observables showed that many intuitive quantities—including global excitation–inhibition balance, assembly overlap, network topology, and temporal statistics of synaptic lifetimes—cannot reliably distinguish one-cycle from multi-cycle realizations. In contrast, a remarkably simple quantity emerged naturally from the snapshot analysis: the cumulative number of \emph{fresh} excitatory neurons recruited during the decay of the residual synaptic memory. This quantity, denoted by $R_E$, predicts the emergence of additional activity cycles with near-perfect accuracy over a broad range of realizations.

Importantly, the predictor is not based on continuing the original network simulation. Instead, it is obtained directly from the synaptic-memory snapshot by evaluating the latent ability of the remaining synaptic configuration to recruit previously available excitatory neurons. In this sense, $R_E$ represents a property of the synaptic state itself rather than a property of the observed neuronal trajectory. The predictor therefore transforms the problem from one of long-term dynamical simulation into one of analyzing a single transient synaptic configuration.

The robustness analysis further strengthens this interpretation. Although increasing the initial activation density substantially changes the total number of neurons that become available for recruitment, the probabilistic relation between $R_E$ and the emergence of additional activity cycles remains almost unchanged over a wide range of initial conditions. This observation suggests that the decisive variable is not the strength of the initial perturbation but the latent excitatory recruitment capacity encoded within the transient synaptic memory.

\subsection{Physical Interpretation and Relation to Existing Memory Theories}
The results obtained in this work suggest a simple physical picture for the regeneration of neuronal activity. The first silent state does not represent the complete disappearance of the memory trace left by the preceding activity cycle. Instead, it corresponds to a transient metastable state \cite{deco2012ongoing, rabinovich2008transient, deco2016metastability, rabinovich2008transient} in which neuronal activity has vanished whereas a substantial fraction of synapses remains active because of their finite lifetimes. Consequently, the network continues to carry information about its recent history despite the absence of active neurons.

From this perspective, the subsequent evolution is governed by the gradual decay of the residual synaptic memory. As inhibitory synapses progressively disappear, neurons that were previously prevented from firing become increasingly excitable. At the same time, the remaining excitatory synapses continue to provide synaptic input inherited from the previous activity episode. Whether these residual excitatory inputs can recruit a sufficiently large population of fresh excitatory neurons before the synaptic traces vanish completely determines whether a new activity cycle will emerge.

This interpretation immediately explains why the predictor identified in the present work performs so well. The quantity $R_E$ is not merely the cumulative number of reactivated neurons. Instead, it measures the latent excitatory recruitment capacity encoded in the transient synaptic-memory state. In other words, it quantifies the ability of the residual synaptic configuration to regenerate collective neuronal activity after complete neuronal silence. The predictor therefore has a clear physical meaning rather than being an empirical quantity obtained solely through numerical observation.

The existence of a narrow overlap region between one-cycle and multi-cycle realizations also becomes naturally understandable within this framework. Far from the transition, the latent recruitment capacity is either clearly insufficient or clearly sufficient to sustain another activity cycle, allowing $R_E$ to classify the two behaviors almost perfectly. Near the transition, however, the network operates close to a critical balance where relatively small differences in the microscopic organization of the residual synaptic currents determine whether the recruited neurons become capable of initiating another collective activation. This explains why the overlap region occupies only a narrow interval of $R_E$ despite the complexity of the underlying dynamics.

These observations also provide an interesting perspective on existing theories of short-term memory. Classical attractor models \cite{amit1989modeling, amit1997model, wang2001synaptic, compte2000synaptic} generally assume that memory is maintained through persistent neuronal firing supported by recurrent excitation. Within this framework, sustained neuronal activity constitutes the primary memory substrate. In contrast, increasing experimental and theoretical evidence \cite{masse2019circuit, mongillo2008synaptic, stokes2015activity, rose2016reactivation, wolff2017dynamic} suggests that information may remain stored during periods of strongly reduced or even completely absent neuronal activity through hidden variables such as transient synaptic modifications.

Our results are consistent with this latter viewpoint. Although all neurons become inactive at the first silent state, the network has not lost its memory. Instead, the residual synaptic configuration completely determines its future evolution and already contains sufficient information to predict whether another activity cycle will occur. The memory variable is therefore not the instantaneous neuronal activity itself but the transient synaptic state inherited from previous activity.

The assembly analysis further reinforces this interpretation \cite{harris2005neural, buzsaki2010neural}. The Jaccard similarity demonstrates that assemblies formed during later activity cycles exhibit only weak overlap with the initial assembly $S_0$. Consequently, repeated activity does not arise from repeatedly activating the same neuronal population. Instead, the active assembly continuously evolves throughout the dynamics. At the same time, the similarity with the assembly formed near the peak of activity remains substantially larger, indicating that the network gradually converges toward dynamically evolving assemblies rather than repeatedly reconstructing its initial state. This behavior is consistent with the general concept of dynamic coding \cite{stroud2024computational, spaak2017stable, murray2017stable}, in which memory is represented by continuously reorganizing neuronal populations instead of a fixed ensemble of persistently active neurons.

Within this dynamic picture, the distinction introduced here between ghost and fresh neurons provides a simple mechanistic interpretation of assembly evolution. Ghost neurons preserve the residual synaptic memory generated by previous activity but cannot establish new outgoing synaptic influence. Consequently, they contribute to memory preservation without promoting network expansion. Fresh neurons, in contrast, constitute the only population capable of generating new synaptic activation and recruiting additional neurons into the evolving assembly. Activity regeneration therefore depends not on repeatedly activating previously active neurons but on continuously recruiting fresh excitatory neurons from the latent synaptic-memory state.

Taken together, these observations suggest that transient synaptic memory and dynamic coding should not be regarded as competing explanations of short-term memory \cite{mongillo2008synaptic, stokes2015activity}. Rather, they describe complementary aspects of the same phenomenon. Transient synaptic memory preserves the latent dynamical state of the network, whereas dynamic coding describes how this hidden memory is expressed through continuously evolving neuronal assemblies. From this perspective, memory is not stored in a fixed population of active neurons, but in the network's latent capacity to recruit new assemblies from the residual synaptic state.

\subsection{Limitations}
Although the proposed framework successfully reveals the mechanism underlying activity regeneration in networks with transient synaptic memory, several limitations should be acknowledged.

First, the present study considers a minimal neuronal network model composed of binary neurons connected through random Erd\H{o}s--R\'enyi \cite{erdds1959random} connectivity with fixed excitatory and inhibitory synaptic strengths. This simplified formulation was deliberately adopted to isolate the dynamical role of transient synaptic memory while avoiding additional complexities arising from heterogeneous neuronal properties, adaptive synapses, or structured connectivity. Whether the same predictive principle remains valid in networks with modular organization \cite{sporns2016networks, khanjanianpak2025optimizing}, scale-free connectivity, or biologically realistic neuronal dynamics remains an important question for future investigation.

Second, the predictor proposed here is obtained through systematic analysis of synaptic-memory snapshots rather than from an analytical derivation of the governing equations. Although its predictive performance approaches perfect classification over the investigated parameter range, deriving the critical latent excitatory reactivation capacity directly from the underlying dynamical equations would provide a deeper theoretical understanding of the transition between one-cycle and multi-cycle behavior.

Finally, the present model is entirely deterministic once the initial condition has been specified. Consequently, the emergence of one-cycle or multi-cycle behavior reflects differences in the transient synaptic-memory state rather than stochastic neuronal fluctuations. Extending the proposed framework to stochastic neuronal dynamics constitutes an important direction for future investigation. Moreover, the present predictor is evaluated only at the first silent state. Whether analogous snapshot-based predictors can characterize subsequent silent periods during longer activity sequences remains an open question.

These limitations, however, primarily concern the quantitative details of the predictor rather than the underlying mechanism itself. The specific numerical threshold separating one-cycle and multi-cycle realizations may depend on the neuronal model, network architecture, and synaptic parameters. Nevertheless, the central physical picture emerging from this work—that activity regeneration is governed by the latent excitatory recruitment capacity encoded in transient synaptic memory—is expected to remain applicable to a considerably broader class of neuronal networks. Similarly, the distinction between ghost and fresh neurons reflects a dynamical property of transient synaptic-memory systems and reflects a general dynamical property of systems with transient synaptic memory rather than a specific feature of the minimal model considered here.

\subsection{Future Directions}
The framework developed in this work naturally suggests several promising research directions. An immediate extension is to examine whether the concept of latent excitatory reactivation capacity remains valid in more realistic neuronal models incorporating heterogeneous neuronal properties, synaptic plasticity, transmission delays, and structured network architectures. Such studies will help determine the extent to which the mechanism identified here represents a general principle of transient synaptic-memory networks rather than a property of the present minimal model.

A second direction concerns the development of an analytical description of the transition between one-cycle and multi-cycle dynamics. In the present work, the critical reactivation capacity emerged from systematic numerical analysis of synaptic-memory snapshots. Deriving this transition directly from the governing dynamical equations would provide a quantitative theory for predicting activity regeneration and could reveal how the critical threshold depends on network topology, synaptic time constants, and excitation--inhibition balance.

Perhaps the most intriguing implication of the present work concerns the possibility of controlling network activity using only information contained in the synaptic-memory snapshot. Since the residual synaptic state already determines the future evolution of the network, a natural extension is the development of state-dependent intervention strategies. This observation naturally suggests a state-dependent intervention strategy. If the synaptic-memory snapshot indicates that the latent excitatory reactivation capacity is below the critical level required for activity regeneration, a brief external cue could be delivered specifically during the silent interval to recruit additional fresh excitatory neurons. Unlike conventional stimulation protocols that continuously perturb network activity, such an intervention would exploit the intrinsic computational state of the network and act only when the residual synaptic memory predicts that spontaneous regeneration is unlikely. In this sense, the silent period becomes not merely a passive interval between activity episodes but an optimal window for targeted intervention.

Beyond its computational significance, this perspective may also have broader implications for understanding and supporting short-term memory. Rather than monitoring ongoing neuronal firing alone, future approaches to monitoring and controlling short-term memory may benefit from estimating the hidden synaptic state that persists after neuronal activity has disappeared. More broadly, the snapshot-based framework introduced here suggests that transient synaptic memory is not merely a passive repository of recent activity, but an active dynamical substrate from which the future evolution of neuronal networks can be predicted, manipulated, and potentially controlled.

\section{Conclusion} \label{sec:conclusion}

In this work, we investigated the mechanism underlying the spontaneous regeneration of neuronal activity in a minimal neuronal network with transient synaptic memory. We showed that the first silent state does not mark the end of the network dynamics. Instead, the residual synaptic-memory snapshot already contains sufficient information to determine whether activity will terminate after a single cycle or regenerate additional cycles. Remarkably, these distinct dynamical outcomes emerge in an otherwise homogeneous network, demonstrating that transient synaptic memory alone is sufficient to diversify future network behavior.

By introducing the distinction between \emph{ghost} and \emph{fresh} neurons, we identified the latent excitatory reactivation capacity, quantified by the cumulative number of fresh excitatory neurons, as a simple yet highly accurate predictor of activity regeneration. Unlike conventional approaches that rely on continuing the full network simulation, this predictor is extracted directly from a single synaptic-memory snapshot, providing a transparent mechanistic interpretation of the regeneration process.

More broadly, our results suggest that transient synaptic memory should be viewed not merely as a passive trace of previous activity, but as an active dynamical substrate that determines the future evolution of neuronal networks. We hope that the snapshot-based framework introduced here will stimulate further studies on transient memory, activity regeneration, and state-dependent control of neuronal dynamics in more realistic brain networks.
\section{Declaration of interests}
The authors declare no competing interests.
\section{Author contribution}
The research project was initiated by Alireza Valizadeh, who formulated the original scientific question and supervised the study. Mozhgan Khanjanianpak developed the model, carried out the theoretical analysis, implemented the simulations, analyzed the data, interpreted the results, prepared figures, and wrote the initial manuscript. Alireza Valizadeh provided scientific guidance throughout the project and critically reviewed and revised the manuscript. Both authors discussed the results and approved the final manuscript.

\section{Data and Code Availability}
\label{sec:material}
The C++ simulation code, Python analysis notebooks, and representative example datasets used to reproduce the figures in this work are publicly available at
\begin{center}
\url{https://github.com/MozhganKhanjanianpak/TransientSynapticMemory}
\end{center}

\bibliography{MyBibFile.bib}

\appendix
\label{sec:appendix}
\setcounter{figure}{0}
\renewcommand{\thefigure}{S\arabic{figure}}
\section*{Appendix}

\begin{figure*}[t!]
	\centering    \includegraphics[width=0.9\linewidth]{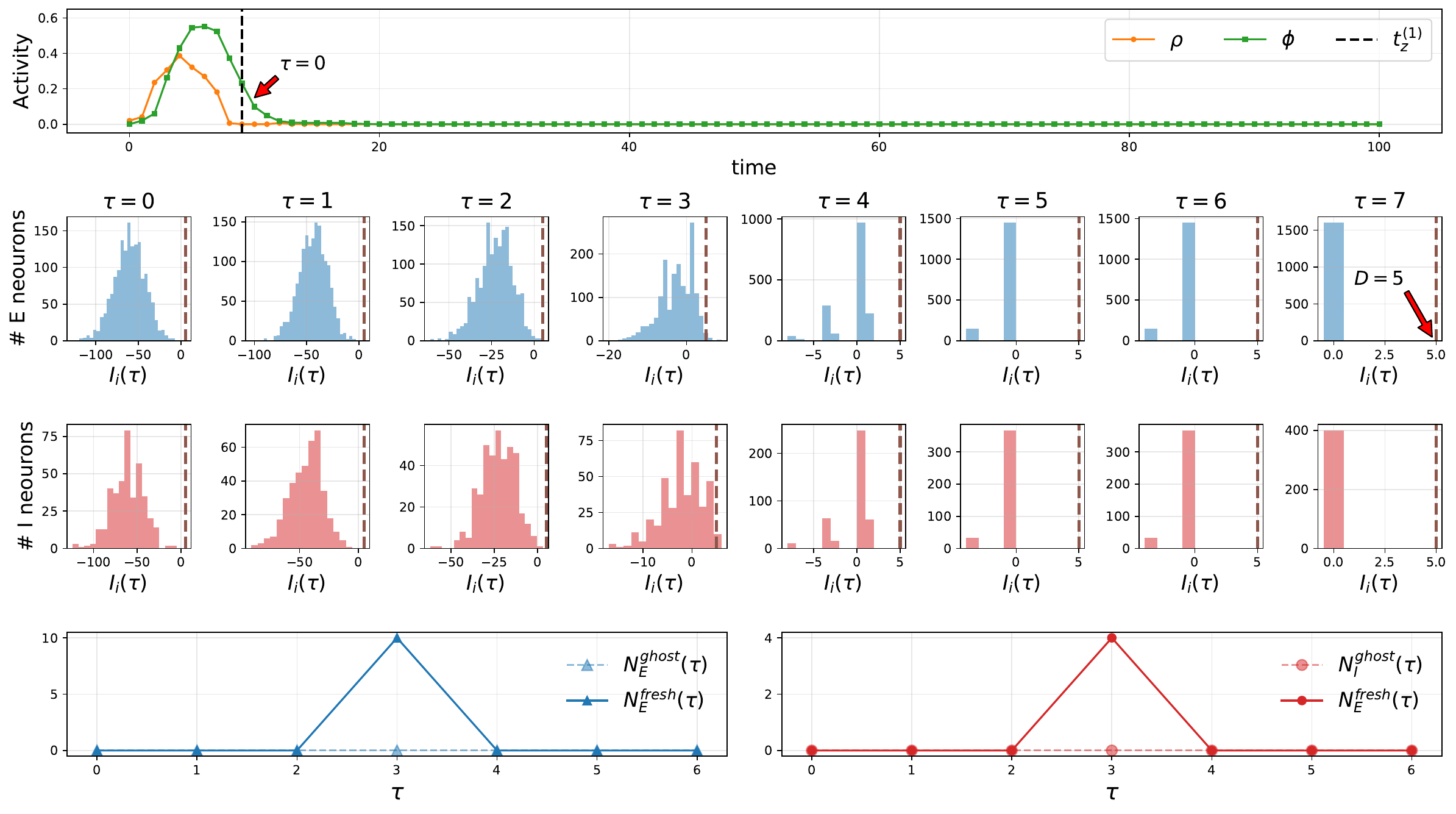}
	\caption{\textbf{One-cycle example with $R_E=10$.}}
	\label{fig:HistogramOvrelapOne}
\end{figure*}

\begin{figure*}[t!]
	\centering    \includegraphics[width=0.9\linewidth]{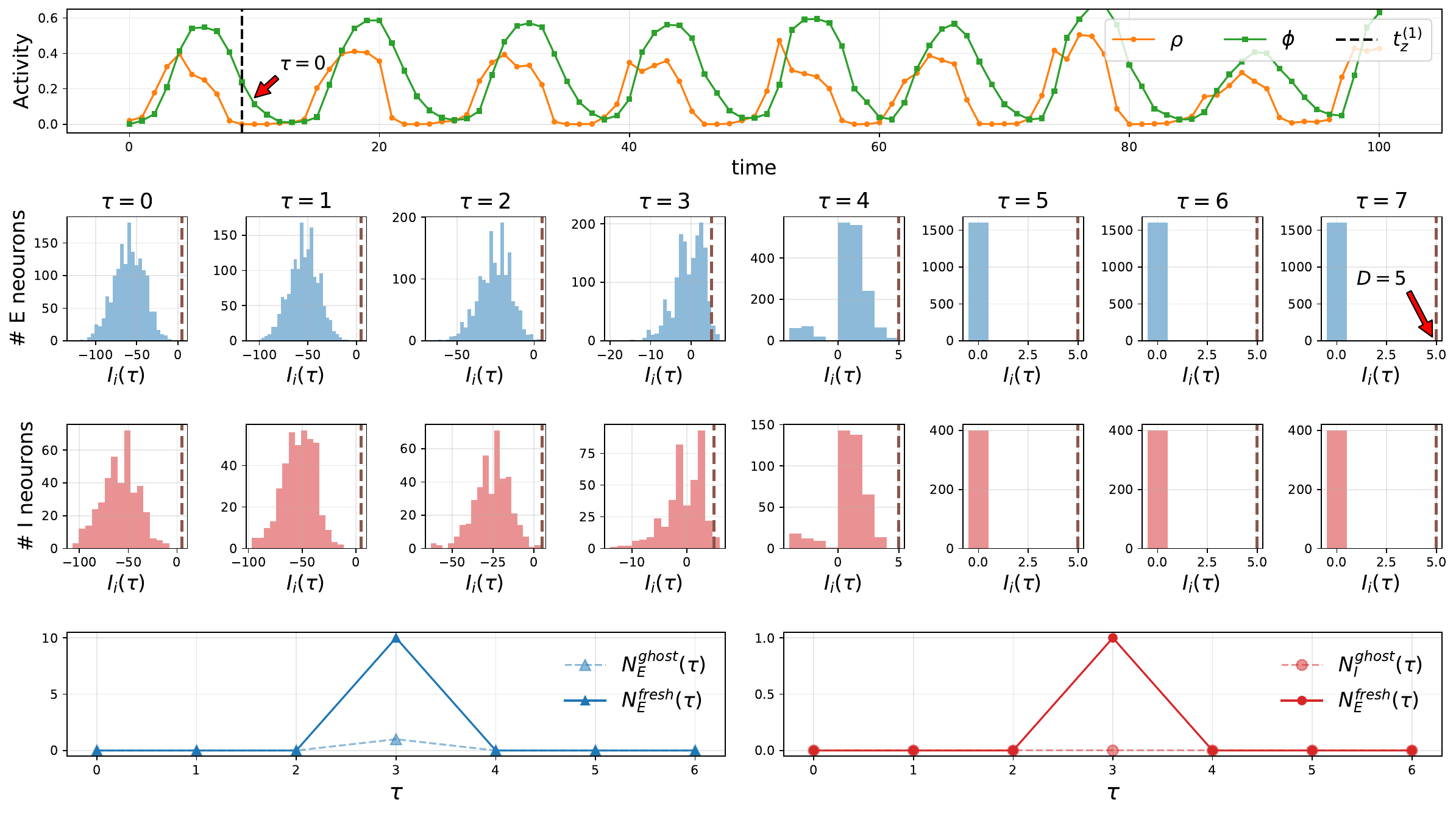}
	\caption{\textbf{Multi-cycle example with $R_E=10$.}}
	\label{fig:HistogramOverlapMulti}
\end{figure*}

\end{document}